\documentclass[prd,preprint,nofootinbib]{revtex4}

\usepackage{graphicx}
\usepackage{amssymb}
\usepackage{amsmath}
\usepackage{color}

\begin{document}

\newcommand{\beq}{\begin{equation}}   
\newcommand{\eeq}{\end{equation}}
\newcommand{\bea}{\begin{eqnarray}}   
\newcommand{\eea}{\end{eqnarray}}
\newcommand{\bear}{\begin{array}}  
\newcommand {\eear}{\end{array}}
\newcommand{\bef}{\begin{figure}}  
\newcommand {\eef}{\end{figure}}
\newcommand{\bec}{\begin{center}}  
\newcommand {\eec}{\end{center}}
\newcommand{\non}{\nonumber}  
\newcommand {\eqn}[1]{\beq {#1}\eeq}
\newcommand{\la}{\left\langle}  
\newcommand{\ra}{\right\rangle}
\newcommand{\ds}{\displaystyle}
\def\SEC#1{Sec.~\ref{#1}}
\def\FIG#1{Fig.~\ref{#1}}
\def\EQ#1{Eq.~(\ref{#1})}
\def\EQS#1{Eqs.~(\ref{#1})}
\def\GEV#1{10^{#1}{\rm\,GeV}}
\def\MEV#1{10^{#1}{\rm\,MeV}}
\def\KEV#1{10^{#1}{\rm\,keV}}
\def\lrf#1#2{ \left(\frac{#1}{#2}\right)}
\def\lrfp#1#2#3{ \left(\frac{#1}{#2} \right)^{#3}}
\newcommand{\red}{\textcolor{red}}

\begin{flushright}
IPMU 10-0074
\end{flushright}

\title{
Probing Variant Axion Models at LHC
}

\author{Chuan-Ren Chen$^{(a)}$,
Paul H. Frampton$^{(a,b)}$, Fuminobu Takahashi$^{(a)}$ and  Tsutomu T. Yanagida$^{(a,c)}$}

\affiliation{${}^{(a)}$Institute for the Physics and Mathematics of the Universe,
University of Tokyo, Chiba 277-8583, Japan\\
 ${}^{(b)}$Department of Physics and Astronomy, University of North Carolina, Chapel Hill, NC 27599-3255, USA \\
${}^{(c)}$Department of Physics,
University of Tokyo, Tokyo 113-0033, Japan
}

\date{\today}

\begin{abstract}
We study collider implications of variant axion models which naturally
avoid the cosmological domain wall problem.  We find that in such
models the branching ratio of $h \rightarrow \gamma\gamma$ can be
enhanced by a factor of $5$ up to $30$ 
as compared with the standard model prediction.  
The $h \rightarrow \gamma\gamma$ process is therefore a
promising channel to discover a light Higgs boson  at the LHC and to
probe the Peccei-Quinn charge assignment of the standard model fields from Yukawa interactions.
\end{abstract}


\maketitle

\section{Introduction}
\label{sec:1}
The strong CP problem is one of the profound problems in the standard
model (SM).  One of the elegant solutions was proposed by Peccei and
Quinn~\cite{Peccei:1977hh}.  They introduced a global chiral
U(1)$_{\rm PQ}$ symmetry, which is explicitly broken by the quantum
chromodynamics (QCD) anomaly. In association with spontaneous breaking
of the PQ symmetry, the axion appears as the pseudo Nambu-Goldstone
boson coupled to the QCD anomaly. The effective CP phase $\theta$ is
set to be $0$ due to the dynamics of the axion.

In order to accommodate the PQ mechanism, we need to introduce an
additional Higgs field and assign appropriate PQ charges to the Higgs
field(s) and the quarks.  However the PQ charge assignment is not
unique, and there is a variety of axion models.  
In addition to the Dine-Fischler-Srednicki-Zhitnitsky
(DFSZ)~\cite{Dine:1981rt,Zhitnitsky:1980tq} and
Kim-Shifman-Vainshtein-Zakharov
(KSVZ)~\cite{Kim:1979if,Shifman:1979if}  invisible axion models, 
 there is a class of models known as variant axion models in which quarks of the same
chirality are assigned different PQ charges and therefore coupled to
different Higgs fields~\cite{Peccei:1986pn,Krauss:1986wx}.  Although
the original motivation of the model was to revive the
Peccei-Quinn-Weinberg-Wilczek (PQWW)~\cite{Peccei:1977hh,
  Weinberg:1977ma, Wilczek:1977pj} axion model by making it consistent
with experiments, it is straightforward to extend the setup to the
invisible axion model.

There is a variety of astrophysical and cosmological constraints on
the axion models~\cite{Raffelt:1996wa,
  Kolb:1990vq,Preskill:1982cy,Kawasaki:2008sn}.  Of particular
importance is the constraint from the cosmological domain wall
problem~\cite{Sikivie:1982qv}.  Suppose that the PQ symmetry is
restored during or after inflation. Then domain walls may be formed in
association of the spontaneous breaking, if there are multiple
degenerate vacua.  Let us denote the multiplicity by $N_{\rm DW}$. The
DFSZ model, which is a natural extension of the original PQWW model,
is plagued with the domain wall problem with $N_{\rm DW} =
3$~\footnote{The value of $N_{\rm DW}$ can be $6$, depending on the
  interaction between the Higgs doublets and the PQ
  singlet~\cite{Geng:1990dv}. }.  On the other hand, the KSVZ model
has $N_{\rm DW} = 1$ and therefore avoids the problem {\it iff} there
is only one heavy quark which carries a PQ charge. Interestingly, the
variant axion models in which one of the two Higgs doublets  couples
to only one quark flavor has $N_{\rm DW} = 1$ and therefore naturally avoid
the domain wall problem~\cite{Geng:1990dv, Hindmarsh:1997ac}.

In this paper we study the variant axion
models~\cite{Peccei:1986pn,Krauss:1986wx} and focus on the
implications for the Higgs boson search at the Large Hadron Collider
(LHC).  In the SM, the dominant production mechanism for a Higgs boson
at the LHC is so-called gluon-gluon fusion (GGF) via a top quark loop,
followed by the vector-boson-fusion (VBF) process through the annihilation of two vector bosons.
If the light Higgs
boson is mainly contained in the scalar field with a PQ charge which
has a larger expectation value, we find that the GGF process can be
highly suppressed or slightly enhanced depending on the quark flavor
to which the Higgs is coupled, and that the VBF process remains
similar to the SM result. 
 Moreover, the decay branching
ratio of the Higgs boson to two photons, $h\to\gamma\gamma$,
which is the most important search channel of the light Higgs boson
at the LHC, can be significantly enhanced due to the suppression in the $h \rightarrow b {\bar b}$ as compared with the SM prediction.

The rest of the paper is organized as follows.  In
Sec.~\ref{sec:model} we briefly explain the variant axion models.  The
phenomenology of the Higgs boson is studied in Sec.~\ref{sec:pheno},
and finally, Sec.~\ref{sec:conclusion} is devoted for discussions and
conclusions.

\section{Variant axion models}
\label{sec:model}
We introduce two Higgs doublets $\Phi_1$ and $\Phi_2$ 
which carry the same quantum numbers of the SM gauge group:
\beq
\Phi_1 = \left(
\begin{array}{c}
\phi_1^+ \\
\frac{1}{\sqrt{2}} (v_1+h_1+i g_1)
\end{array}
\right),~~~~
\Phi_2 = \left(
\begin{array}{c}
\phi_2^+ \\
\frac{1}{\sqrt{2}} (v_2+h_2+i g_2)
\end{array}
\right),
\label{twohiggs}
\eeq
where $v_1$ and $v_2$ are the vacuum expectation values (vev).  For
later use, we define $\tan \beta \equiv v_2/v_1$ with $0 \leq \beta
\leq \pi/2$.  We assign PQ charges $0$ and $-1$ to $\Phi_1$ and
$\Phi_2$, respectively~\footnote{ This choice is for understanding the
  Yukawa structure. In order to get rid of the mixing with the
  Nambu-Goldstone boson eaten by the $Z$ boson, we need to assign the PQ charges of opposite sign and equal
  magnitude  to both Higgs fields.  }.  A PQ singlet field $\sigma$ which carries
a PQ charge $1$ is assumed to have a large vev $v (\,\geq \GEV{9})$ in
order to make the axion invisible~\cite{Raffelt:1996wa}. 
Based on the quantum numbers of the Higgs and PQ scalar fields, we have the following renormalizable scalar potential, 
\bea
V(\Phi_1, \Phi_2, \sigma) &=& \lambda_1 \left(|\Phi_1|^2 - \frac{v_1^2}{2} \right)^2
+\lambda_2 \left(|\Phi_2|^2 - \frac{v_2^2}{2} \right)^2
+\lambda \left(|\sigma|^2 - \frac{v^2}{2} \right)^2\non\\
&& + a\, |\Phi_1|^2 |\sigma|^2 + b\,|\Phi_2|^2  |\sigma|^2+ \left(m\, \Phi_1^\dag \Phi_2 \sigma + {\rm h.c.}\right)
\non\\
&&+ d\, |\Phi_1^\dag \Phi_2 |^2 + e \left|\Phi_1 \right|^2  \left|\Phi_2 \right|^2.
\label{eq:potential}
\eea

In order to avoid the domain wall problem, we couple $\Phi_2$ to only
one quark flavor, while $\Phi_1$ is coupled to the rest of the quarks
and the leptons~\footnote{ We can couple $\Phi_2$ instead of $\Phi_1$ to the
  leptons without affecting the domain wall problem. Then, e.g. $h
  \rightarrow \tau^+\tau^-$ would be enhanced (or suppressed).} .
  We consider the following three cases; $\Phi_2$ is
coupled to (1) the $u$ quark; (2) the $c$ quark; and (3) the $t$ quark
by assigning a PQ charge $-1$ to $u_R$, $c_R$, or $t_R$ in each
case~\footnote{ If the weak mixing angles come
  from only a rotation of the up-quark sector, we can safely assign the PQ charge $-1$ to
  $d_R$, $s_R$ or $b_R$ instead of $u_R$, $c_R$ or $t_R$.
  }.  The relevant Yukawa interactions are given by
\beq
-{\cal L}_{\rm Yukawa}  \;=\; y_{ij}^{(d)} {\bar Q}_{L i} \Phi_1 d_{Rj}+
y_{i}^{(u)} {\bar Q_{L i}} \tilde{\Phi}_3 u_{R} + y_{i}^{(c)} {\bar Q_{L i}} \tilde{\Phi}_4 c_{R}
+y_{i}^{(t)} {\bar Q}_{L i} \tilde{\Phi}_5 t_{R},
\label{eq:yukawa}
\eeq
where the subscripts $i$ and $j$ denote the generations, and $Q_L$,
$u_R$, $d_R$ are the left-handed quark doublet, right-handed up-type
and down-type quarks, respectively, and $\tilde{\Phi}_a \equiv i
\sigma_2 \Phi_a^*$. The three models correspond to
\bea
{\rm Model~U} &:& \Phi_3 = \Phi_2,~~\Phi_4=\Phi_1,~~\Phi_5=\Phi_1,\non\\
{\rm Model~C} &:& \Phi_3 = \Phi_1,~~\Phi_4=\Phi_2,~~\Phi_5=\Phi_1,\non\\
{\rm Model~T} &:& \Phi_3 = \Phi_1,~~\Phi_4=\Phi_1,~~\Phi_5=\Phi_2.
\label{eq:uct}
\eea

The above potential (\ref{eq:potential}) contains a massless degrees
of freedom, the axion $a$, which mainly resides in the phase of
$\sigma$. The axion acquires a coupling to the QCD anomaly term
through the Yukawa interactions (\ref{eq:yukawa}), and thus solving
the strong CP problem~\cite{Peccei:1986pn,Krauss:1986wx}.

Note that, since we couple different Higgs fields to the quarks of the
same chirality, the flavor violation is not automatically
absent~\cite{Glashow:1976nt}. However the flavor constraint does not
affect the following discussion, and so, we simply neglect the
tree-level Higgs-mediated flavor changing processes in the following
discussion.

The peculiar Yukawa structure of the variant axion model makes it
cosmologically viable since the domain wall problem is
absent. Interestingly, as we will see in the next section, due to the
Yukawa structure, the production and decay branching ratios of the
light Higgs boson are significantly modified.

\section{Phenomenology}
\label{sec:pheno}
After the electroweak symmetry is broken, two CP even Higgs bosons are
left at the weak scale.  The heavy and light Higgs fields, denoted by
$H$ and $h$, respectively, are mixtures of $h_1$ and $h_2$ defined in Eq.~(\ref{twohiggs}) and can be
written as
\beq
\left(
\begin{array}{c}
H\\
h
\end{array}
\right) \;=\; \left(
\begin{array}{cc}
\cos \alpha & \sin \alpha \\
-\sin\alpha & \cos \alpha
\end{array}
\right)
\left(
\begin{array}{c}
h_1\\
h_2
\end{array}
\right),
\eeq
where $\alpha$ is a mixing angle varying from $-\pi/2$ to $\pi/2$.  In
the limit of $\sin \alpha \rightarrow 0$, the light (heavy) Higgs
mainly contains $h_2$ ($h_1$), and bear in mind that $\Phi_2$ is
assigned a PQ charge.

The Higgs couplings to the gauge bosons are given by
\bea
HVV &:& \cos(\beta-\alpha)\, \,g_{\rm SM}^{hVV},\\
hVV &:& \sin(\beta-\alpha)\,\, g_{\rm SM}^{hVV},
\eea
where $V$ denotes either $W$-boson or $Z$-boson, and $g^{hVV}_{\rm SM}$ denotes
the coupling between the corresponding gauge bosons to the Higgs boson
in the SM.  Therefore, if we consider a small mixing angle, i.e.,
$|\sin \alpha| \ll 1$, with moderately large $\tan \beta \gtrsim 5$,
the couplings of the light Higgs boson to the gauge bosons are almost the
same as the SM case~\footnote{ In this limit, the light Higgs boson
  $h$ mainly resides in the Higgs field with a larger vev, $\Phi_2$. }.  We
take $\tan\beta = 5$ throughout this paper as a reference value in the
numerical results.  The difference among the three models is therefore the 
couplings of the light Higgs to heavy quarks.  As can be seen from
Eqs. (\ref{eq:yukawa}) and (\ref{eq:uct}), $\Phi_1$ generates all the
fermion masses except for the $u$ ($c, t$) quark in the model U (C,
T), resulting in different branching ratios and production cross
sections for the light Higgs boson.  We will focus on the light Higgs
boson $h$ whose mass is smaller than $130$ GeV and its decay to two
photons, $h\to\gamma\gamma$, since this  is the main search channel at
the LHC.  We assume all the other degrees of freedom (the heavy
neutral, CP-odd, and charged Higgs bosons) are heavy enough to be
neglected in the following study.\footnote{
For instance, there is a lower bound on the charged Higgs boson mass $M_{H^+} \gtrsim 300$\,GeV~\cite{Hewett:1996ct}
in the model T.}

\subsection{Model U}
\label{sec:model_u}
In this setup, the $u$ quark is special since it is coupled to
$\Phi_2$, while the other fermions are coupled to $\Phi_1$.
The couplings of the light Higgs boson to the up, charm, bottom and
top quarks are given by
\bea
huu&:& \frac{\cos \alpha}{\sin \beta} g_{\rm SM}^{huu},\\
hcc &:&- \frac{\sin \alpha}{\cos \beta} g_{\rm SM}^{hcc},\\
hbb &:&- \frac{\sin \alpha}{\cos \beta} g_{\rm SM}^{hbb},\\
htt &:& -\frac{\sin \alpha}{\cos \beta} g_{\rm SM}^{htt},
\eea
where $g_{\rm SM}^{hff}$ is the coupling of 
fermion $f$ to the Higgs boson in the SM.
The other down-type quarks and leptons have couplings
similar to the bottom quark case. We can drop the $huu$ coupling in the following analysis, since the
$u$ quark Yukawa coupling is tiny, and it does not play any important
role at collider experiments.

We are interested in the case of $|\sin \alpha| \ll 1$ and $\tan \beta
\gtrsim 5$, for which the light Higgs boson $h$ is contained mainly in
$h_2$~\footnote{In order for the top Yukawa coupling not to blow up at high energies,
$\tan \beta$ cannot be much larger than unity in the models U and C. For a smaller value of $\tan \beta (\gtrsim 1)$,
the production cross section of the Higgs through VBF and the decay width 
of $h \rightarrow \gamma\gamma$ become smaller. Note however that the branching fraction of $h \rightarrow \gamma \gamma$ is 
not sensitive to $\tan \beta$ and can be still enhanced over the SM value, as long as $h \rightarrow W^+W^-$ is the dominant decay mode.
}.  The couplings to the fermions other than the $u$ quark are
then enhanced by $\tan \beta$, but suppressed by $\sin \alpha$. If the
mixing $\sin \alpha$ is sufficiently small (i.e., $|\sin \alpha| \ll \cot \beta$), the couplings to those
fermions, especially to the heavy quarks, can be suppressed.  If this
is the case, the partial decay width of the light Higgs to a fermion
pair (except for the $u$ quark) will be highly suppressed compared to
the SM prediction, while $W^+W^-$ and $ZZ$ decay modes remain almost
unchanged.  Since the $h\to gg$ is mainly through a top-quark loop,
its partial width is suppressed as well.  For $h\to\gamma\gamma$, the
partial decay width is slightly enhanced; this can be understood from
the fact that the contribution from the top-quark loop, which
partially cancels that of the $W$-boson loop in the SM, is now
suppressed.  Note that the main decay mode of the Higgs is $h
\rightarrow b {\bar b}$ for $M_h \lesssim 130$\,GeV in the
SM. Therefore, if the $h \rightarrow b {\bar b}$ is suppressed, the
branching fractions of all the other modes are enhanced accordingly.

\begin{figure}[h]
\includegraphics[scale=0.6]{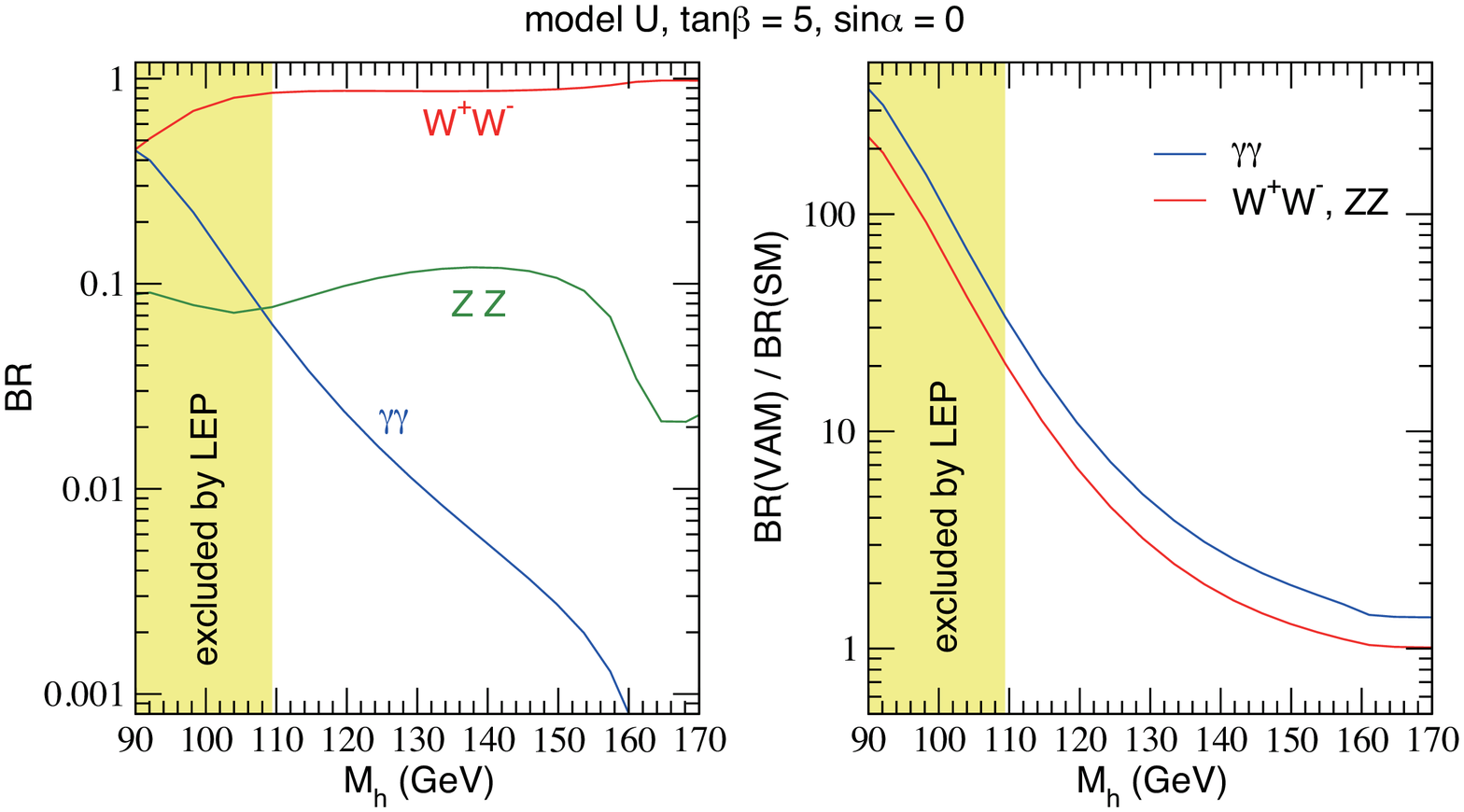}
\includegraphics[scale=0.6]{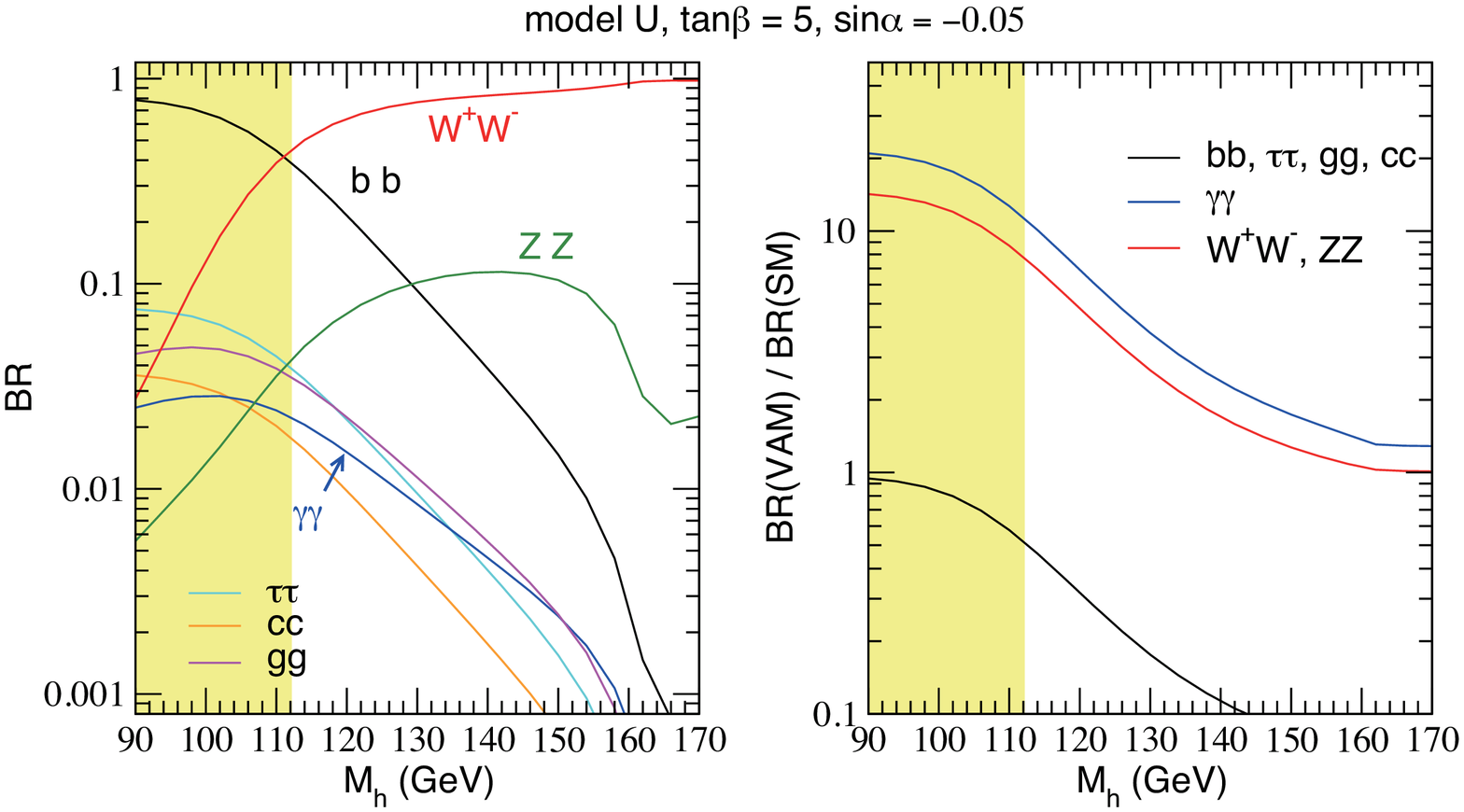}
\caption{
The branching fractions of various decay processes of the light Higgs 
for the model U, as a function of the Higgs mass $M_h$ (left). The ratios to the SM predictions
are also shown (right). (VAM stands for the variant axion model.) The upper panels are for the limiting case of $\sin \alpha = 0$,
and the bottom ones are for $\sin \alpha = -0.05$.
}
\label{fig:u}
\end{figure}

We show the branching fractions of various decay processes and the
enhancement or the suppression factor compared to the SM prediction as
a function of $M_h$ in Fig.~\ref{fig:u}. We consider two cases: no
mixing ($\sin\alpha = 0$) in the upper panel and a small (but
non-zero) mixing ($\sin\alpha = -0.05$) in the lower
panel. Furthermore, due to the absence of the signals of the Higgs
boson in the past and current experiments at LEP and Tevatron, the
mass and the branching ratios of the Higgs boson are constrained, as
shown by the shaded regions in Fig.~\ref{fig:u}~\footnote{The decay
  branching ratios of the light Higgs boson are calculated by
  modifying the HDECAY~\cite{Djouadi:1997yw} and the constrains are
  obtained using HiggsBounds~\cite{Bechtle:2008jh}. }.  In the case of
$\sin\alpha = 0$, the Higgs boson with a mass smaller than about $109$
GeV is excluded by the direct search of $h\to\gamma\gamma$ at the LEP
II~\cite{lepgamma}.  The decay branching ratio of $h\to\gamma\gamma$
can be as large as $1\%$ to $6\%$ for a Higgs boson lighter than $130$
GeV, which is about a factor of $5$ to $30$ enhancement compared to
the SM prediction. If the mixing angle $\alpha$ is small but non-zero,
the $h\to b\bar{b}$ decay branching ratio can become
non-negligible. The $h\to\gamma\gamma$ is enhanced by a factor $4$ at
$M_h = 130$ GeV and by a factor of $10$ at $M_h = 112$ GeV below which
the mass region is excluded by the search of $h\to b\bar{b}$ in LEP
II~\cite{smhiggs}.

\subsection{Model C}
\label{sec:model_c}
\begin{figure}[t]
\includegraphics[scale=0.6]{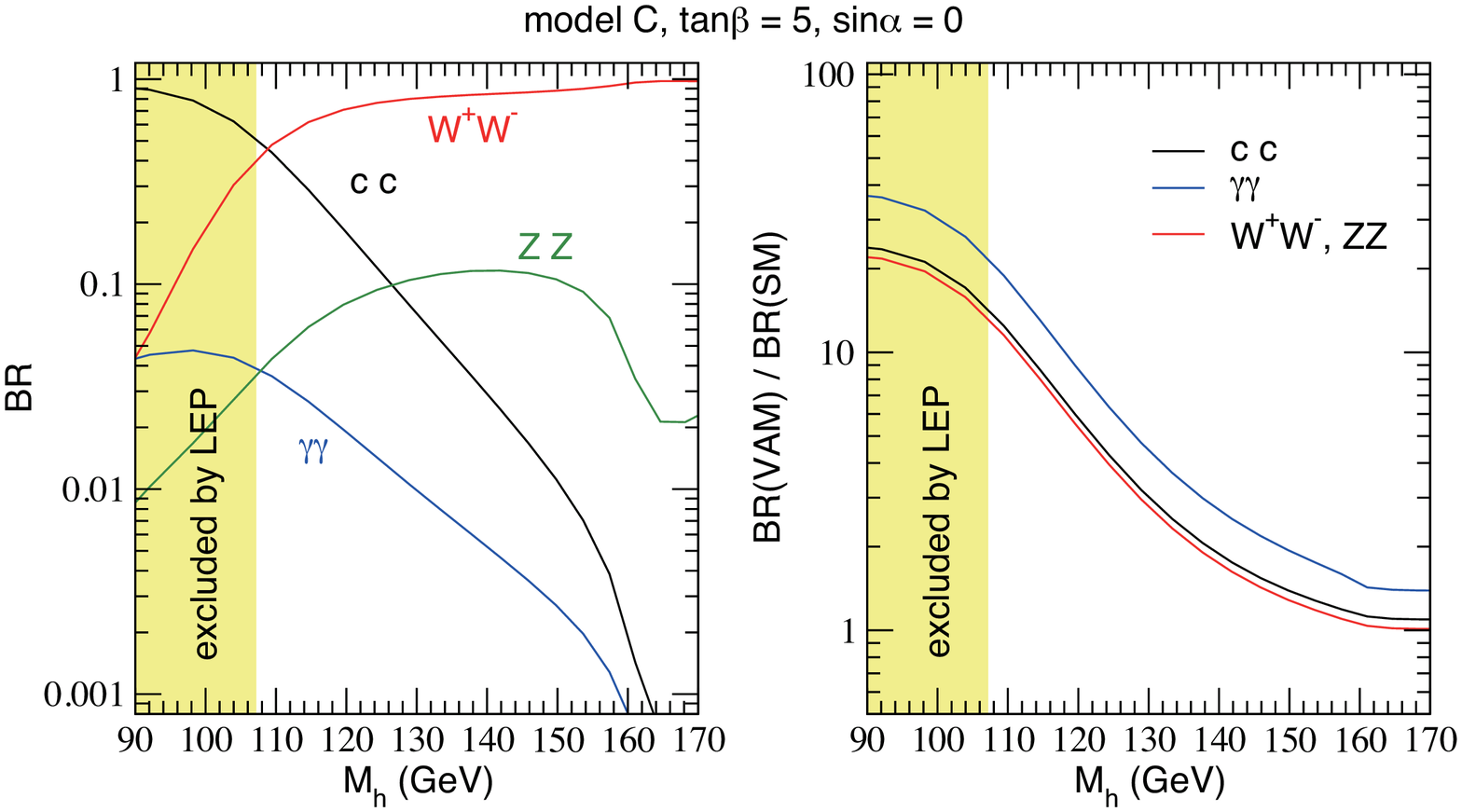}
\includegraphics[scale=0.6]{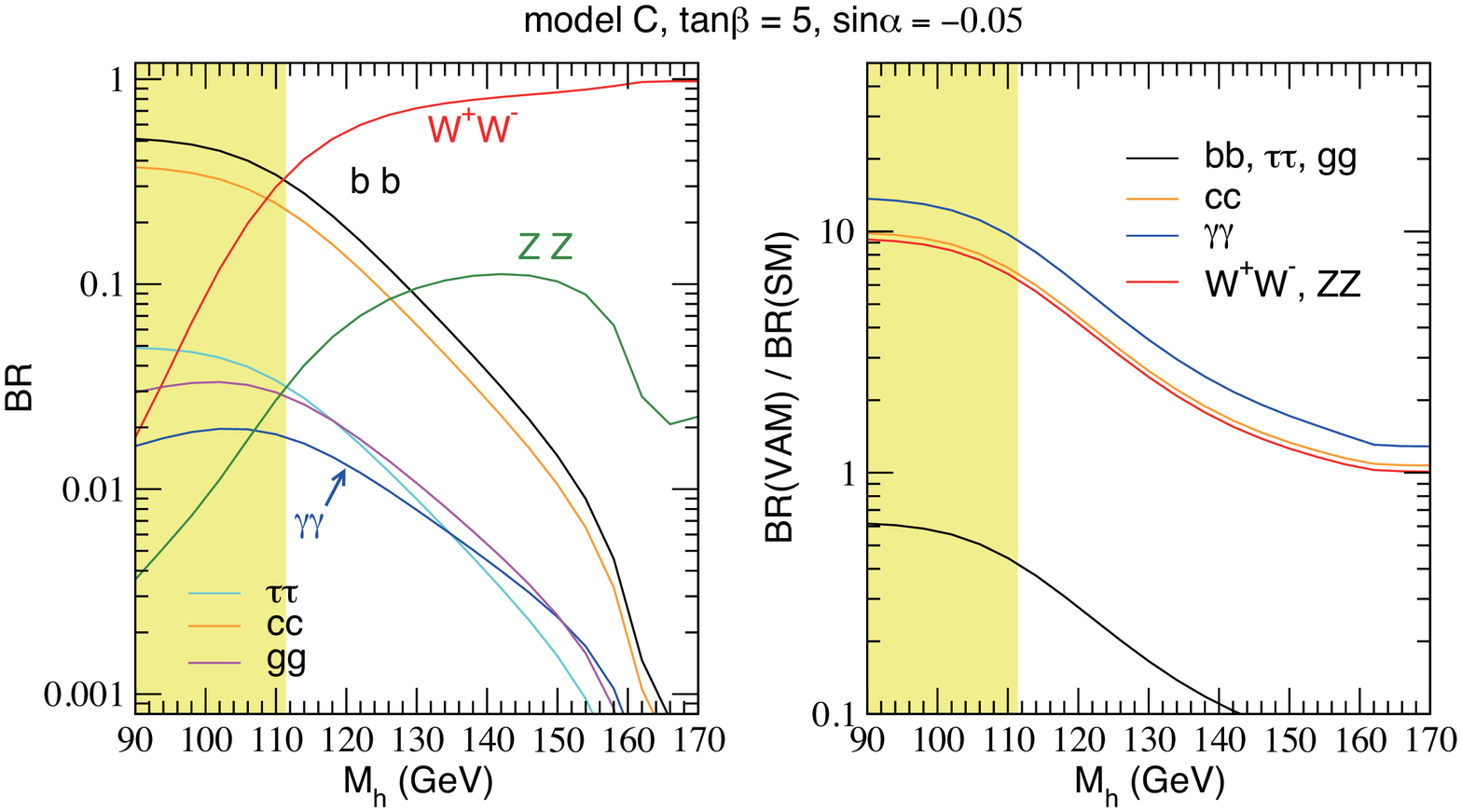}
\caption{Same as Fig.~\ref{fig:u} but for the model C. }
\label{fig:c}
\end{figure}

The couplings of the light  Higgs to the charm, bottom and top quarks are given by
\bea 
hcc&:& \frac{\cos \alpha}{\sin \beta} g_{\rm SM}^{hcc},\\ hbb
&:&- \frac{\sin \alpha}{\cos \beta} g_{\rm SM}^{hbb},\\ htt &:&
-\frac{\sin \alpha}{\cos \beta} g_{\rm SM}^{htt}, 
\eea 
where the notations follow the description in the model U.  The main
difference from the model U is that the decay into a charm quark pair
is not suppressed even in the limit of $|\sin \alpha| \ll \cot \beta$ and $\tan
\beta \gtrsim 5$, since the $hcc$ coupling approaches to the SM value.
Because of the existence of large $h\to c\bar{c}$ decay branching
ratio, the enhancement of $h\to\gamma\gamma$ is not as large as the
model U at $M_h \simeq 110$ GeV. As $M_h$ increases, the branching
fraction of $h\to\gamma\gamma$ becomes similar to that in model U,
since the $h\to W^+W^-$ decay mode quickly dominates the total decay
rate. As we can see from Fig.~\ref{fig:c}, when $M_h$ is smaller than
$130$ GeV, the branching fraction of $h \rightarrow \gamma \gamma$ can
be as large as a few percent and about $1\%$ for $\sin\alpha = 0$ and
$\sin\alpha = -0.05$, respectively. The excluded mass region in the
upper (lower) panel of Fig.~\ref{fig:c} is from direct search of
$h\to\gamma\gamma$ ($h\to b\bar{b}$) in the LEP
II~\cite{lepgamma,smhiggs}.

\subsection{Model T}
\label{sec:model_t}
The couplings of the light  Higgs to the  charm, bottom and top quarks are given by
\bea
hcc &:&- \frac{\sin \alpha}{\cos \beta} g_{\rm SM}^{hcc},\\
hbb &:&- \frac{\sin \alpha}{\cos \beta} g_{\rm SM}^{hbb},\\
htt &:& \frac{\cos \alpha}{\sin \beta} g_{\rm SM}^{htt}.
\eea
The main difference from the other two models is that the $htt$
coupling is not suppressed.  Therefore the effective higgs-gluon-gluon
coupling is not suppressed, which enhances the decay branching ratio
of Higgs to two gluons, $h\to g g$, compared to the other two models. We should mention several
advantages in such a setup. The Yukawa structure with large $\tan
\beta$ explains why the top quark is much heavier than the other
quarks. Furthermore, the flavor changing neutral process does not give
any constraints; the top flavor physics in a similar setup has been
extensively studied in Ref.~\cite{Baum:2008qm}.

\begin{figure}[t]
\includegraphics[scale=0.6]{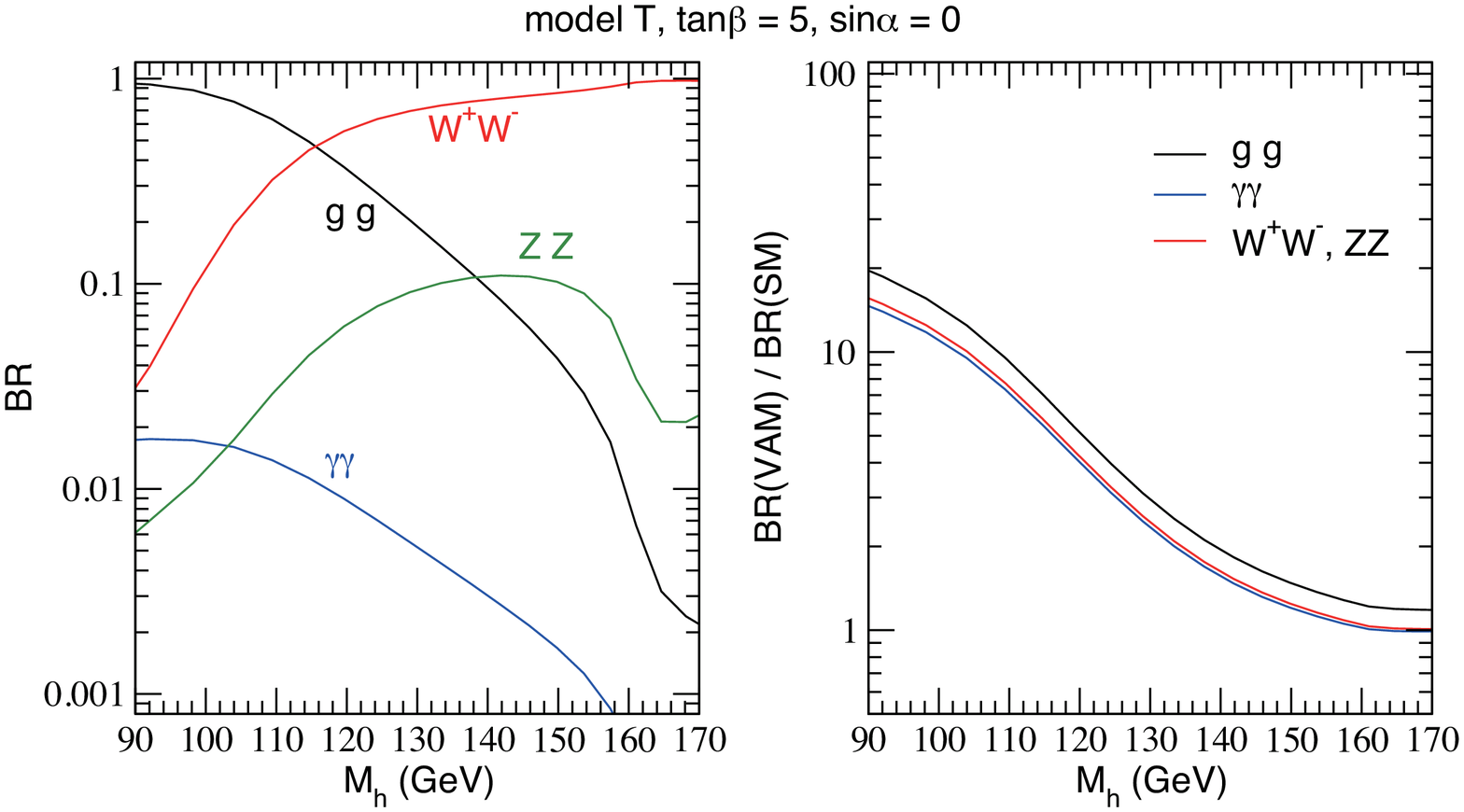}
\includegraphics[scale=0.6]{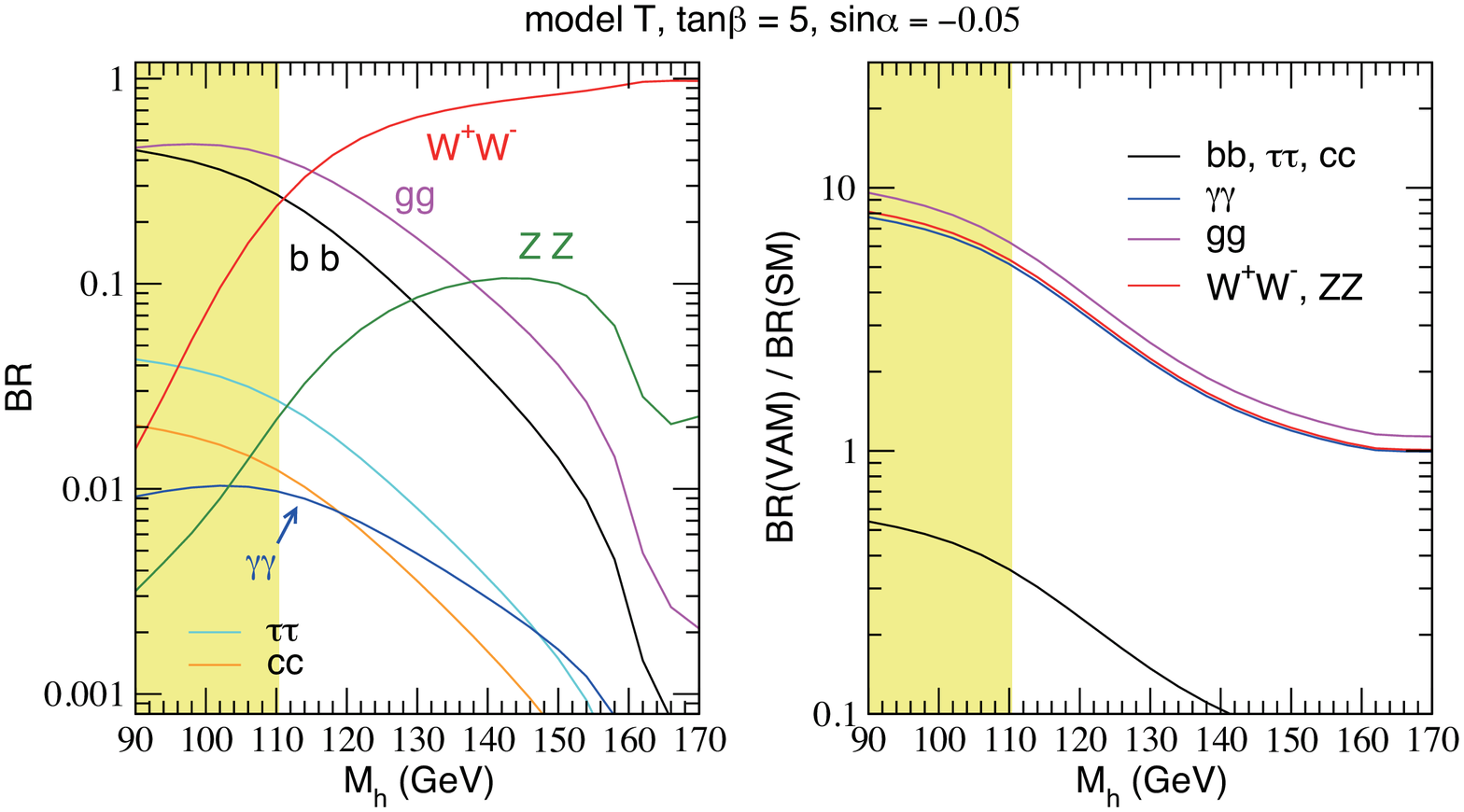}
\caption{Same as Fig.~\ref{fig:u} but for the model T.}
\label{fig:t}
\end{figure}

Since the branching ratio of $h\to\gamma\gamma$ is not as large as the
model U and model C, the result of LEP II does not impose any
constrain for $M_h$ larger than $90$ GeV when $\sin\alpha = 0$, as
shown in upper panel of Fig.~\ref{fig:t}.  For $\sin\alpha = -0.05$,
similarly, $h\to b\bar{b}$ becomes sizable, and the LEP II
result~\cite{smhiggs} excludes the Higgs boson lighter than about
$110$ GeV.

\subsection{Production cross sections and Higgs search at the LHC}
\label{sec:production}
\begin{figure}[t]
\includegraphics[scale=0.6]{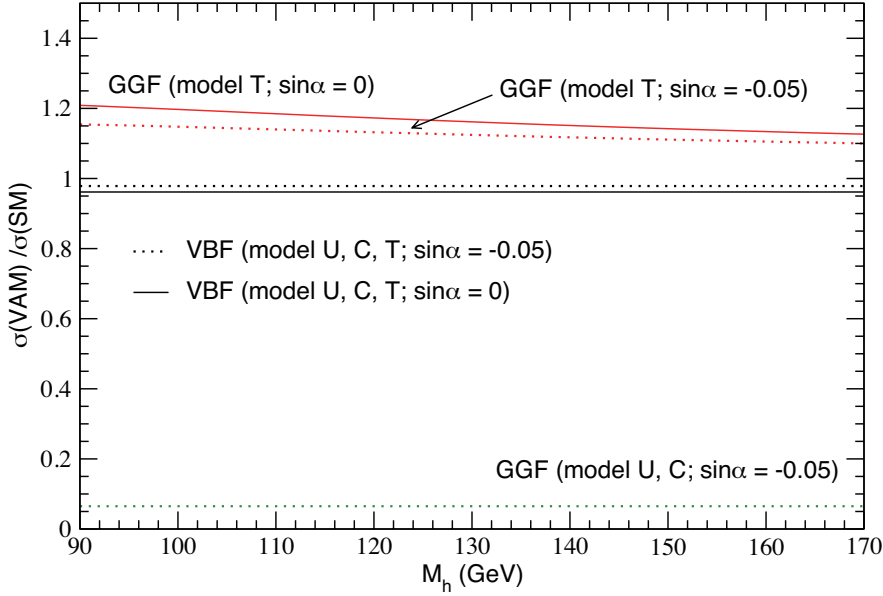}
\caption{The ratio of the production cross sections to the SM values for the
gluon-gluon fusion (GGF) and the vector-boson fusion (VBF) processes,
as a function of the light Higgs mass, $M_h$. The solid lines are for the case of $\sin \alpha = 0$,
while the dotted lines are for $\sin \alpha = -0.05$.}
\label{fig:production}
\end{figure}

In this section we discuss the prospect for the discovery of the light Higgs for each model and 
we assume $|\sin \alpha| \ll \cot \beta$ and $\tan \beta \gtrsim 5$ in the following discussion
unless otherwise stated.
As we mentioned previously, in the SM, the dominant production of a Higgs boson
at the LHC is GGF while VBF is sub-dominant.  In the models
U and C, since the coupling of the Higgs boson to the top quarks is suppressed, the
GGF production cross section will decrease significantly and VBF is
the same as the SM result. Therefore, the VBF may become the leading
production mechanism for the light Higgs boson.  On the other hand, in
the model T, there is no suppression in the coupling of the Higgs
boson to top quarks and gauge bosons, so the production cross section
is similar to that in the SM. The production cross sections of GGF and
VBF compared to the SM for each model are shown in
Fig.~\ref{fig:production}.  For $\sin\alpha=0$, the GGF process is
completely turned off in the models U and C, which is not shown in the
plot. In the model T, the GGF process can be enhanced by about $20\%$ for
$M_h\lesssim 130$ GeV (red solid line).

At the LHC, $h\to\gamma\gamma$ is  the main search channel for a light Higgs boson.
It has been shown that the LHC is able to
discover such a signal for the SM Higgs boson lighter than about $130$ GeV,
with the $30$ $fb^{-1}$ integrated luminosity and $14$ TeV center-of-mass (c.m.) energy 
in the inclusive search~\cite{cmstdr,
  atlastdr}. Furthermore, the CMS collaboration~\cite{cmstdr} has also
studied exclusively $h\to\gamma\gamma$ channel in VBF and in the
production of a light Higgs boson in association with a gauge boson,
$q \bar{q}^{\prime} \to W^{\pm*}/Z^{*}\to W^{\pm}/Z h$, and the estimated
significance is about $2.2\sigma$ with $30$ $ fb^{-1}$ luminosity for
both processes with $115$ GeV $\lesssim M_h \lesssim 130$ GeV.

For the variant axion models U and C, the GGF process is significantly
suppressed, therefore the VBF and associated production with vector
boson (VH) become important. For example, 
the branching ratio of $h \rightarrow \gamma\gamma$
is enhanced by a factor of about $8$ ($5$) for $M_h = 120$\,GeV in model C when $\sin\alpha = 0$ ($\sin\alpha = -0.05$).
One can then estimate from the SM study that such a light Higgs boson should be discovered
with only $3$ $fb^{-1}$ ($10$ $fb^{-1}$) luminosity by event counting, assuming that 
the production cross sections for VBF and VH are not changed. In the
case of the model T, the situation will be more improved compared to the model
U and C since the cross sections of all the production processes are
similar to the SM predictions while the branching ratio of
$h\to\gamma\gamma$ is enhanced.
 Comparing with the
inclusive production cross section of the Higgs boson at the designed
$14$ TeV c.m. energy at the LHC, the production cross section will be
reduced by a factor of $3\sim 4$~\footnote{The factor is estimated
  from the decrease of GGF using Higlu~\cite{Spira:1995mt}, since the
  GGF is dominant process and is about ten times larger than the
  sub-leading one.}  in the current operation of LHC with $7$ TeV
c.m. energy. If the background decreases by the same
factor~\footnote{The true suppression factor should be calculated
  precisely for the $7$ TeV c.m. energy at the LHC.}, we expect, by event
counting, a strong evidence of the existence of the light Higgs boson
with $1 fb^{-1}$ integrated luminosity by the end of 2011. For
instance, the statistical significance of the signal can be larger than
$3\sigma$ ($2\sigma$) for a $110$ GeV Higgs boson if $\sin\alpha = 0$
($\sin\alpha=-0.05$).

\section{Discussion and Conclusions}
\label{sec:conclusion}
The variant axion models have a special Yukawa structure, and its
phenomenology is similar to the two Higgs doublet models.  
The model U is similar to the 2HDM type I, in which
one of the Higgs doublets has no Yukawa couplings. Indeed,
for a large $\tan\beta \gg 1$ and $\sin\alpha=0$, the model U is almost same as the
fermiophobic Higgs scenario~\cite{Haber:1978jt,Landsberg:2000ht}, which has been studied by the
LEP~\cite{Rosca:2002me} and Tevatron experiments~\cite{Aaltonen:2009ga}.
It is also possible to shut off the light Higgs decay to the down-type quarks 
in the supersymmetric SM~\cite{Loinaz:1998ph,Carena:1998gk}, which have some similarities
with the model T~\cite{Landsberg:2000ht}. One important difference is that the $h \rightarrow c{\bar c}$
is suppressed in the model T.

The axion is a plausible candidate for dark matter in our model. The
abundance is determined by the breaking scale of the PQ symmetry, $f_a
= v$. The axion dark matter is produced from (1) coherent
oscillations~\cite{Preskill:1982cy}  and (2) axion string and domain wall
decays~\cite{Vilenkin:1982ks,Harari:1987ht,Yamaguchi:1998gx,Lyth:1991bb,Nagasawa:1994qu,Chang:1998tb}. Although
the production from the axionic strings and walls involves a
relatively large uncertainty,  the required value of $f_a$ for the
axion to account for the observed dark matter density is of
$O(10^{10})\sim O(10^{11})$ GeV. If the axion is the main component of dark matter,
the axion direct search (e.g. ADMX~\cite{Asztalos:2009yp}) may be able
to detect it.

In this paper we have studied the collider implications of the variant axion
models which naturally avoid the cosmological domain wall problem.  
We have found that,  if the light Higgs boson contains mainly the neutral Higgs field which
carries a PQ charge, the branching fraction of $h
\rightarrow \gamma \gamma$ can be significantly enhanced for a
moderately large value of $\tan \beta$,  due to the suppression of the
couplings to the heavy fermions. For the models U and C
(T) the enhancement factor can reach $4$ to ${\cal O}(10)$ ($2$ to
${\cal O}(10)$) for $M_h$  lighter than $130$ GeV.  
 Since the GGF process is highly suppressed in the models U and C,
the VBF and VH processes are important.
With the large decay branching ratio of $h\to\gamma\gamma$, the LHC
with $14$ TeV c.m. energy will have a greater potential to discover the light Higgs
boson with a low luminosity ($\lesssim 10$ $ fb^{-1}$).  
In the model T, the production cross sections for the relevant processes are
almost the same as the SM predictions. Therefore it will be much easier to
discover  the Higgs boson through the search of $h\to\gamma\gamma$,
compared to the SM. Even for the early operation of LHC with $7$ TeV c.m. energy
and $1$ $fb^{-1}$ liminosity, we may be able to have a strong evidence of the
existence of the light Higgs boson.

\begin{acknowledgements}
FT thanks M. Kawasaki for discussion. The work of P.H.F. was supported in
part by U.S. Department of Energy Grant No. DE-FG02-05ER41418. The work of FT was 
supported by the Grant-in-Aid for Scientific Research on Innovative Areas (No. 21111006) and Scientific Research
 (A) (No. 22244030), and JSPS Grant-in-Aid for Young Scientists (B)
 (No. 21740160).  This work was supported by World Premier
 International Center Initiative (WPI Program), MEXT, Japan.

\end{acknowledgements}



\end{document}